\documentclass[proc]{revtex4}
\usepackage{graphicx}
\usepackage{fancyhdr}
\usepackage{graphics}
\usepackage{epstopdf}
\usepackage{textpos}

\usepackage{xspace}%

\newcommand{\Ttt}{\ensuremath{\PSQt \to \cPqt \PSGczDo}\xspace}
\newcommand{\TbW}{\ensuremath{\PSQt \to \cPqb \PSGcp}\xspace}

\newcommand{\wjets}{\ensuremath{\PW+\text{jets}}\xspace}

\newcommand{\lsp}{\PSGczDo\xspace}
\newcommand{\chipo}{\ensuremath{\chipm_1}}

\newcommand{\chiOneZero}{\ensuremath{\tilde{\chi}^{0}_{1}}\xspace}
\providecommand{\MT}{\ensuremath{M_{\cmsSymbolFace{T}}}\xspace}

\input{pdefs}

\begin{document}

\title{\centering 3rd generation SUSY searches at CMS}

\author{
\centering
\begin{center}
Michael Sigamani for the CMS Collaboration
\end{center}}
\affiliation{\centering University of Ghent, Belgium}
\begin{abstract}

The latest results on searches for stop and sbottom squarks are presented.
Searches for direct squark production in a variety of decay channels are reviewed.
The results are based on 19.5 fb$^{-1}$ of LHC proton-proton collisions at $\sqrt{s}=8~$\TeV\
taken with the CMS detector.

\end{abstract}

\maketitle
\thispagestyle{fancy}

\section{Introduction}

\begin{figure}[hbt]
  \begin{center}
        \includegraphics[width=0.33\linewidth]{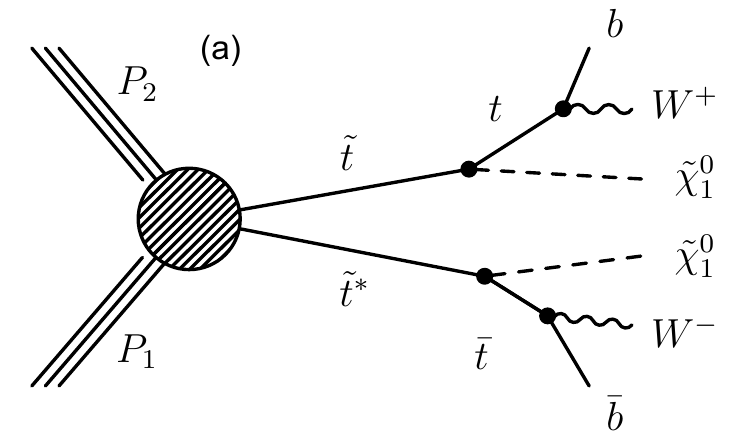}%
        \includegraphics[width=0.33\linewidth]{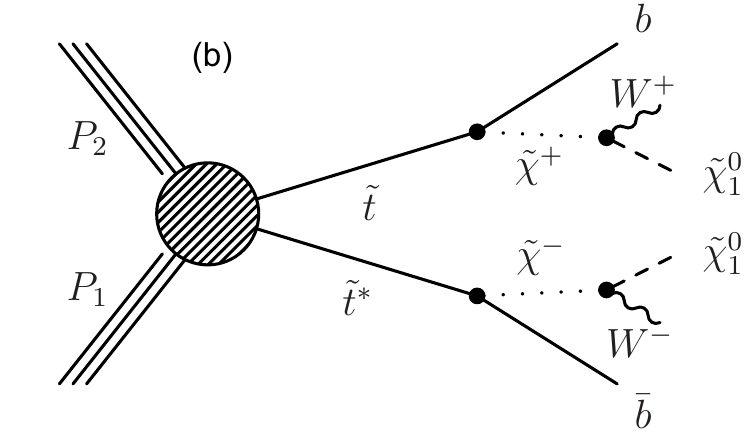}%
  		\includegraphics[width=0.33\linewidth]{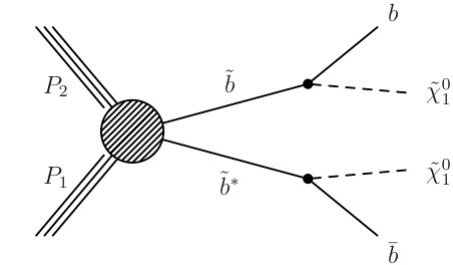}
		\includegraphics[width=0.33\linewidth]{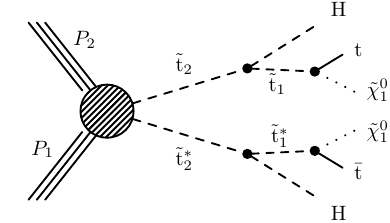}%
		\includegraphics[width=0.33\linewidth]{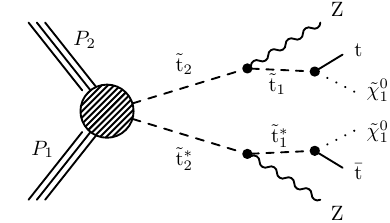}%
		\includegraphics[width=0.33\linewidth]{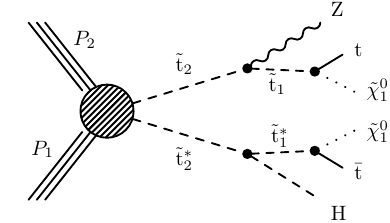}
	\caption{Top row: Diagram for top-squark pair production for (a) the $\Ttt \to \cPqb \PW \chiz_{1}$ decay mode and (b) the $\TbW \to \cPqb \PW \chiz_{1}$ decay mode. 
			 (Right) Bottom-squark pair production. Bottom row: Diagrams for the production of the heavier top-squark ($\sttwo$) pairs followed by the decays $\sttwo\to \PH \stone$ 
             or $\sttwo\to \cPZ \stone$ with $\stone \to \cPqt \PSGczDo$. The symbol * denotes charge conjugation.}
	\label{fig:SigDiagram}
      \end{center}
\end{figure}

The standard model (SM) has been extremely successful at describing particle physics phenomena over the last half-century, 
and the recently discovered boson with a mass of 125~\GeV \cite{CMSHiggs,ATLASHiggs} could be the 
final particle required in this theory, the Higgs boson. 
However, the SM is not without its shortcomings, for instance one requires fine-tuned cancellations of large quantum corrections
in order for the Higgs boson to have a mass at the electroweak
symmetry breaking scale~\cite{SUSY1,SUSY5,SUSY6}. 
This is otherwise known as the hierarchy problem. 
Due to the magnitude of this fine-tuning, one suspects that there is some dynamical mechanism which makes this fine-tuning ``natural''.
Supersymmetry (SUSY) is a popular extension of the SM which postulates
the existence of a sparticle for every SM particle. 
These sparticles have the same quantum numbers as their SM counterparts but differ by one half-unit of spin. 
The loop corrections to the Higgs boson mass due to these sparticles 
are opposite to those of the SM particles thus providing a
natural solution to the hierarchy problem (through the cancellations of the quadratic divergences of the top-quark and top-squark loops).
Furthermore, one expects relatively light top and bottom squarks, with masses below around 1~\TeV,
if SUSY is to be the natural solution to the
hierarchy problem~\cite{Barbieri:1987fn,deCarlos1993320,Dimopoulos1995573,Papucci:2011wy}.
In addition, in $R$-parity conserving SUSY models, the lightest super-symmetric particle (LSP) is often the lightest neutralino \chiOneZero. 
The \chiOneZero offers itself as a good dark matter candidate subsequently 
explaining particular astrophysical observations~\cite{DarkMatterReview,DMGeneral}. 
In this note, several CMS~\cite{Chatrchyan:2008aa} searches are reported for third generation sparticles 
using 19.5 fb$^{-1}$ of LHC proton-proton collisions at $\sqrt{s}=8~$\TeV\, 
of which the Feynman diagrams can be seen in Fig.~\ref{fig:SigDiagram}.

\section{Search for direct stop pair production} 

Searches for top-squark pair production have been performed by
the ATLAS Collaboration at the LHC in several final states~\cite{ATLAS1,ATLAS2,ATLAS3,ATLAS4},
and by the CDF~\cite{CDFstop} and \DZERO~\cite{D0stop} Collaborations at the Tevatron.
In this section two CMS searches using the one-lepton and zero-lepton final states will be discussed.

\subsection{One-lepton final state} 

For this search two particular decay modes of the top squark (\PSQt) namely \Ttt\ and \TbW are studied.
These modes are expected to have large branching fractions if
kinematically allowed. Here $\cPqt$ and $\cPqb$ are the top and bottom quarks, and
the neutralinos (\chiz) and charginos (\chipm) are the mass eigenstates
formed by the linear combination of the gauginos and higgsinos, which
are the fermionic
superpartners of the gauge and Higgs bosons, respectively.
The charginos are unstable and may subsequently decay into neutralinos and \PW\ bosons, leading to the following processes
of interest: $\Pp\Pp\to\PSQt\PSQt^{*}\to \ttbar\chiz_1\chiz_1 \to \bbbar\PW^+\PW^-\chiz_1\chiz_1$
and $\Pp\Pp\to\PSQt\PSQt^{*}\to \bbbar\chip_1\chim_1 \to \bbbar\PW^+\PW^-\chiz_1\chiz_1$, as displayed in Fig.~\ref{fig:SigDiagram} (top left and top middle).
The lightest neutralino $\chiz_1$ is considered to be the stable LSP,
which escapes without detection.

Candidate signal events are required to contain one isolated
lepton (e or $\mu$), no additional isolated track or hadronic $\tau$-lepton candidate, at least four jets with at least one
b-tagged jet, and $\MET > 100\GeV$; this is referred to as the ``preselection''.
Signal regions are defined
demanding $\MT>120\GeV$.
This requirement provides large suppression of the SM backgrounds while retaining high signal efficiency.
Requirements on several kinematic quantities
or on the output of boosted decision tree (BDT) multivariate discriminants
are also used to define the signal regions, as described below.
The analysis is based on events where one of the \PW\ bosons decays leptonically and the other hadronically. This
results in one isolated lepton and four jets, two of which originate
from b quarks. The two neutralinos and the neutrino from the \PW\
decay can result in large missing transverse momentum (\MET).
The major SM background contributions in this search arise from
events with a top-antitop (\ttbar) quark pair where one top quark
decays hadronically and the other leptonically, and from events with a
$\PW$ boson produced in association with jets (\wjets).
These backgrounds, like the signal, contain a single leptonically
decaying \PW\ boson.  The transverse mass, defined as
$\MT \equiv \sqrt{\smash[b]{2 \MET \pt^{\ell} (1-\cos(\Delta\phi))}}$, where $\pt^{\ell}$ is the transverse momentum of the
lepton and $\Delta\phi$ is the difference in azimuthal
angles between the lepton and \MET directions,
has a kinematic endpoint $\MT\ <M_\PW$ for these backgrounds, where $M_\PW$ is the \PW\ boson mass.
For signal events, the presence of LSPs in the final state allows \MT\ to exceed $M_\PW$. 
The dominant background with large \MT\
arises from the ``dilepton \ttbar" channel,
\ie, \ttbar\ events where both \PW\ bosons decay leptonically but
with one of the leptons not identified.
In these events the presence of two neutrinos can lead to large values
of \MET\ and \MT.

\begin{figure}[hbt]
  \begin{center}
    \includegraphics[width=0.45\textwidth]{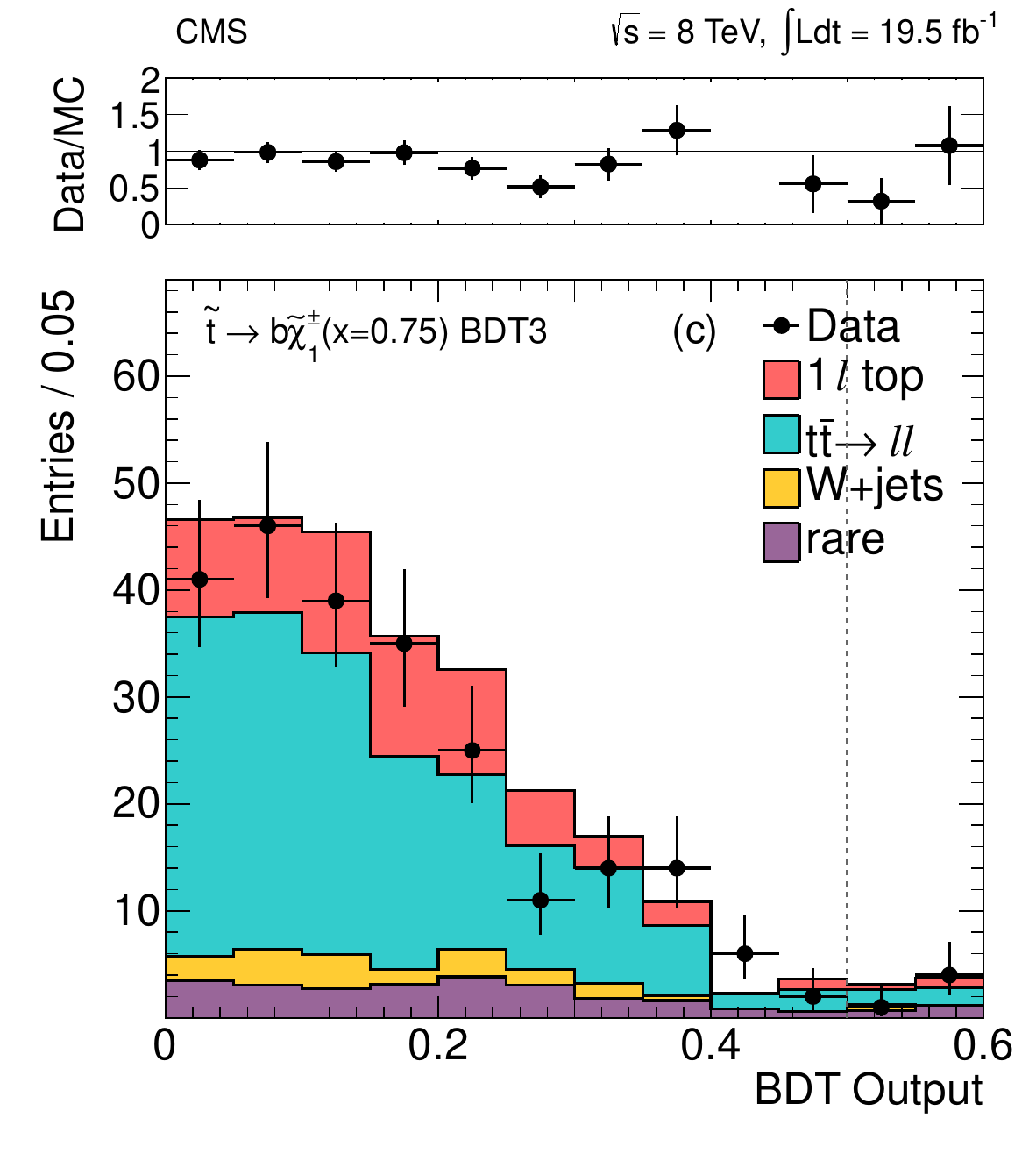}%
    \includegraphics[width=0.45\textwidth]{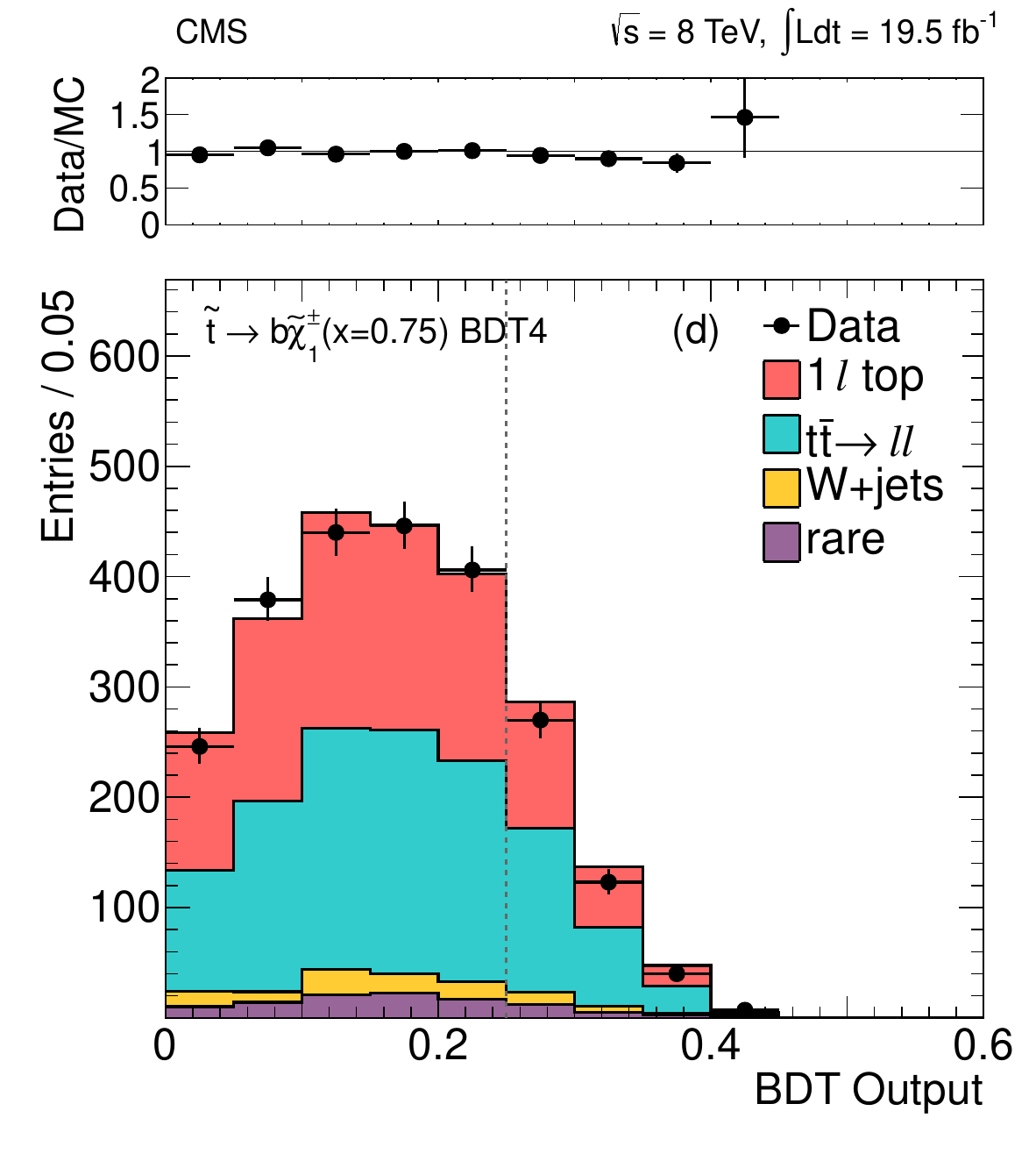} \\
	\caption{
      Comparison of data and MC simulation for the distributions of BDT output and \MT\ corresponding
      to the $x=0.75$ \TbW\ scenario in two particular training regions.  
      \label{fig:plots7}}
      \end{center}
\end{figure}

To define the signal regions a BDT multivariate approach implemented via the {\sc TMVA} package~\cite{TMVA}. 
The preselection requirements are applied 
and the BDT is used to exploit differences of correlation between signal and SM expectations 
for various kinematic variables. 
Separate BDTs are trained for the $\Ttt$ and $\TbW$ decay modes and for
different regions of parameter space.
Example BDT outputs after the preselection are shown in
Fig.~\ref{fig:plots7} for two of the
sixteen BDTs. One notes that the data are in agreement with the MC simulation of SM processes. 
The full details of the event selection with analysis procedure can be found in Ref.~\cite{Chatrchyan:2013xna}.

\begin{figure*}[tbhp]
\centering
\includegraphics[width=0.45\textwidth]{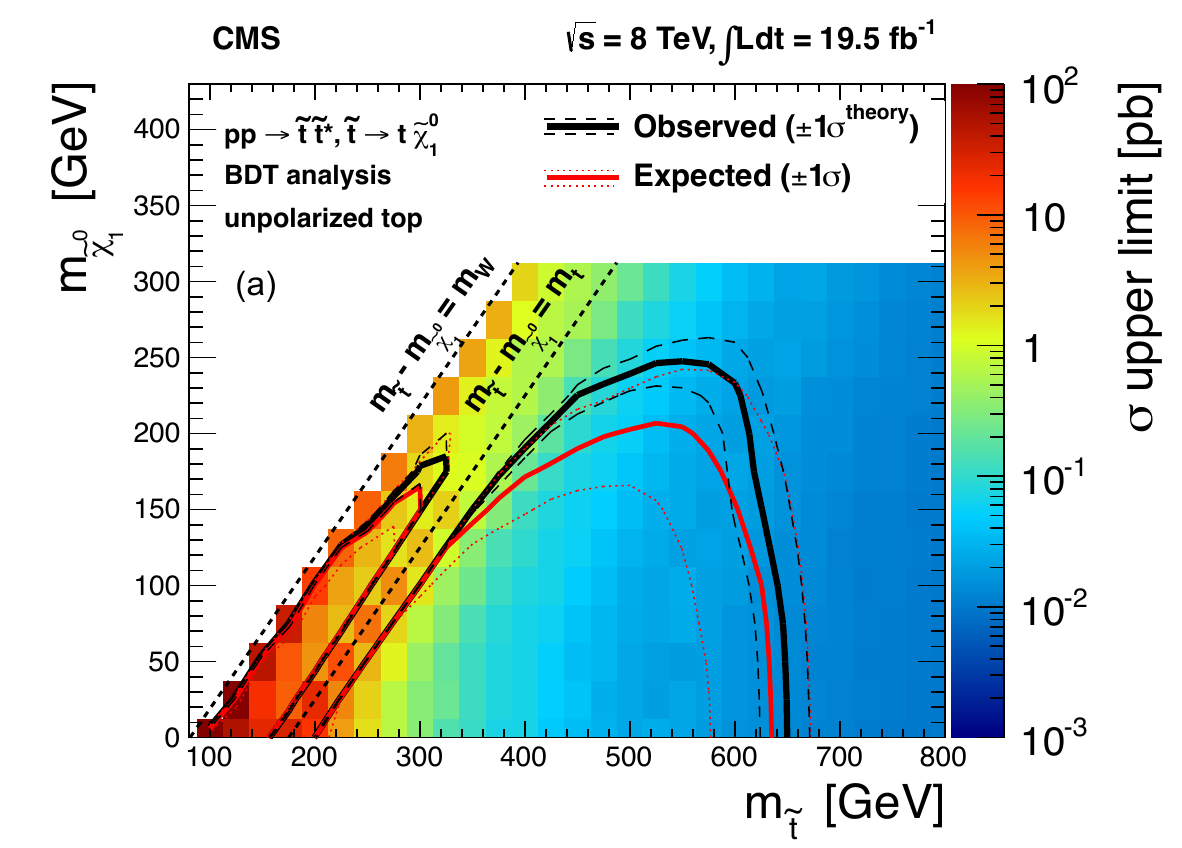} %
\includegraphics[width=0.45\textwidth]{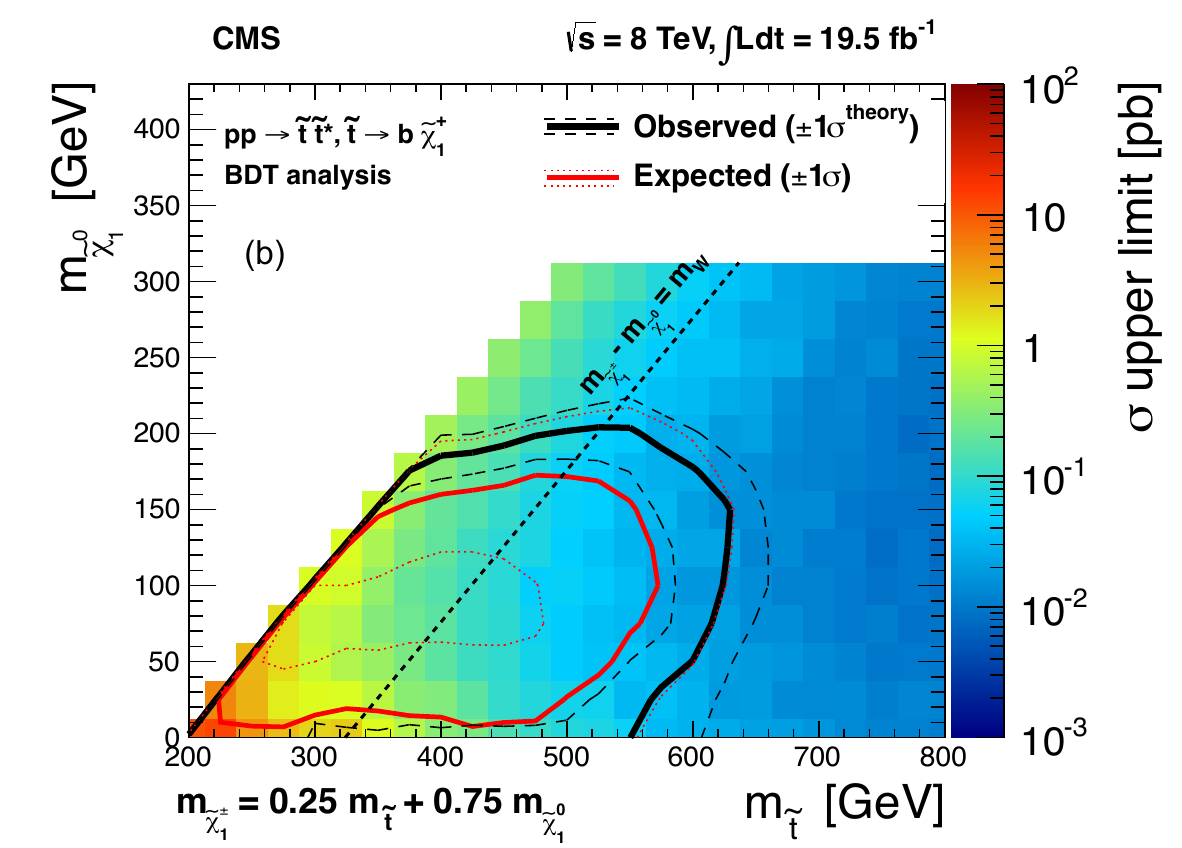}
\includegraphics[width=0.45\textwidth]{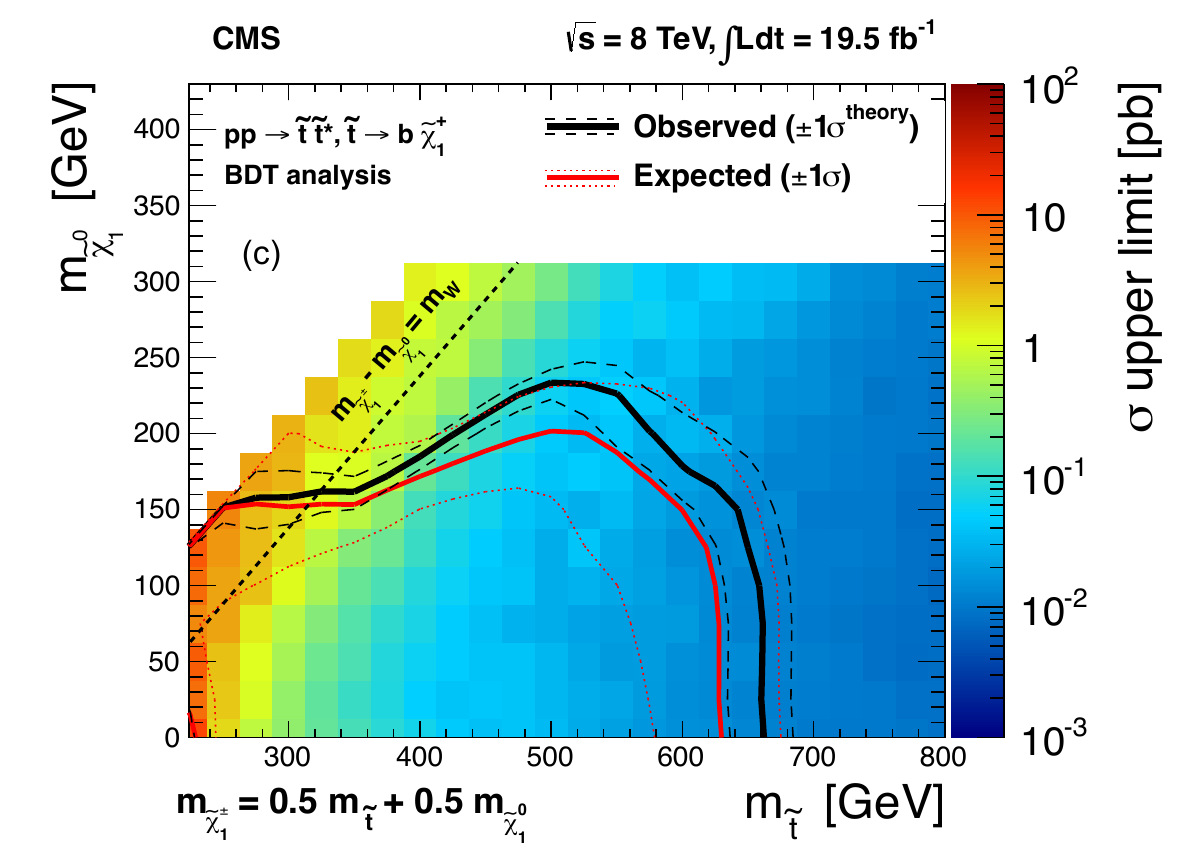} %
\includegraphics[width=0.45\textwidth]{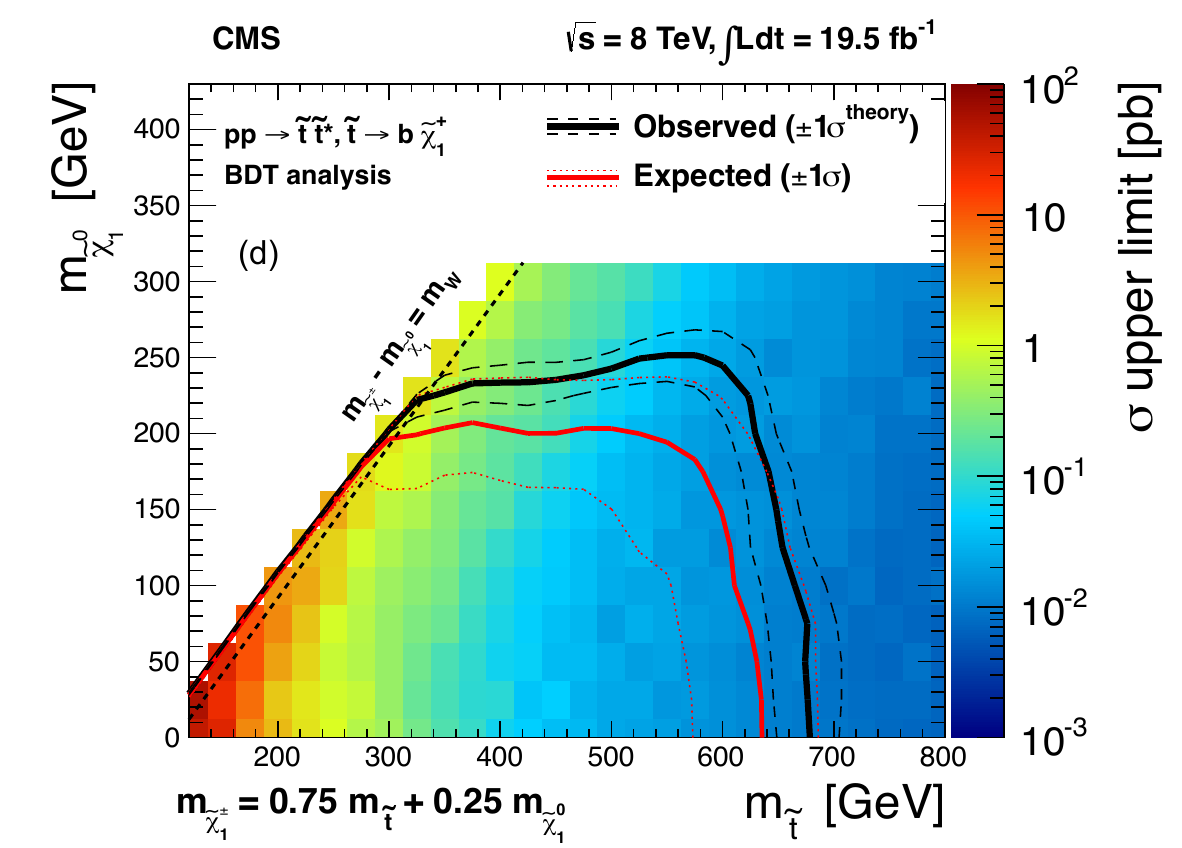}
\caption{
Interpretations using the primary results from the BDT method.
(a) \Ttt\ model; 
(b) \TbW\ model with $x=0.25$;
(c) \TbW\ model with $x=0.50$;
(d) \TbW\ model with $x=0.75$;
The color scale indicates the observed cross section upper limit.
The observed, median expected, and $\pm1$ standard deviation ($\sigma$)
expected 95\% CL exclusion contours are indicated.
The variations in the excluded region due to ${\pm}1\sigma$ uncertainty
of the theoretical prediction of the cross section for top-squark pair
production are also indicated.
\label{fig:bdt_interpretations}
}
\end{figure*}

Overall the observed data yields are consistent with SM expectations and
subsequently limits are placed on the signal model of scalar top quark pair production.
Our results probe top squarks with masses between approximately 150 and 650\GeV,
for neutralinos with masses up to approximately 250\GeV, depending on the details of the model.
For the \Ttt\ search, the results are not sensitive to the model points with $m_{\PSQt}-m_{\lsp}=M_{\text{top}}$ because
the \lsp\ is produced at rest in the top-quark
rest frame. However the results are sensitive to scenarios with
$m_{\PSQt}-m_{\lsp}<M_{\text{top}}$
in which the top quark in the decay \Ttt\ is off-shell, including
regions of parameter space
with the top squark lighter than the top quark.
Furthermore, the acceptance depends on the polarization of
the top quarks in the \Ttt\ scenario, and on the polarization of
the charginos and \PW\ bosons in the \TbW\ scenario. These
polarizations depend on the left/right mixing of
the top squarks and on the mixing matrices of the neutralino and chargino~\cite{polarization1,polarization2}.
The limits on the top-squark and \lsp\ masses
vary by $\pm$10--20\GeV depending on the top-quark polarization.
For the \TbW\ scenario, the acceptance depends on the polarization of the chargino, and on whether the \PW\lsp\chipo\ coupling is
left-handed or right-handed.
In the nominal interpretations for the \TbW\ models presented in Fig.~\ref{fig:bdt_interpretations}, the signal events
are generated with an unpolarized chargino and a left/right-symmetric \PW\lsp\chipo\ coupling.

\subsection{All hadronic final state}

Here the \Ttt decay mode of the top squark (Fig.~\ref{fig:SigDiagram} top left) is studied. 
The analysis is based on events where both of the \PW\ bosons decays hadronically. 
Candidate signal events are selected by requiring five or more jets, no reconstructed e or $\mu$,
at least one jet identified as a b-quark jet, large $\MET$, 
a reconstructed top quark, and several topological requirements. 

The major SM background contributions in this search arise from 
pairs of top quarks, \ttbar, with one of the W bosons from the top quarks
decaying into a neutrino and a lepton,
\Z{}\,+\,jets with the \Z boson decaying into a pair of neutrinos,
and W+jets with the W decaying into a neutrino and a lepton.
Events containing a leptonic decay of a W boson still pass the search selection
criteria if the e or $\mu$ escapes detection or a $\tau$ decays hadronically.

Four search regions are defined by requiring $\MET>200$\GeVc and $\MET>350\GeVc$ with at least 1 or at least 2 b-tagged jets.
The requirement of $\mathrm{N_{b-jets}}\ge2$ increases the sensitivity for high mass top squark production.
The observed data yields are consistent with SM expectations and 
subsequently limits are placed on the signal model of scalar top quark pair production.
As the four search regions are not mutually exclusive, one of the four search regions
is selected at each point in the signal topology scan based on the best expected upper limit
for providing the resulting cross section upper limit. 
The full details of the event selection with analysis procedure can be found in Ref.~\cite{CMS:2013nia}.
The observed cross section upper limits on the signal model considered are shown in
Fig.~\ref{fig:limitsT2ttT2bb} (left).

\section{Search for direct sbottom pair production}

Here the search focusses on SUSY particles produced in proton-proton
collisions with a large imbalance in transverse momentum and two energetic jets one or both of which are
identified as originating from bottom quarks (b-jets). 
This final state is motivated by the production
of pairs of bottom squarks (\sBot) where each \sBot decays into a bottom quark
and a \chiOneZero (see Fig.~\ref{fig:SigDiagram} top right). 
Candidate signal events are selected by requiring 
two central jets with \pt{}$>$ 70 \GeV and $|\eta|<$ 2.4, 
one or both of the leading jets are required to be b-jets, 
no reconstructed e or $\mu$,
\HT{}$>$250 \GeV, \MET{}$>$175 \GeV,
and several topological requirements.
To suppress SM processes such as $\ttbar$ and $\wlnubr$+jets, the invariant
transverse mass of the sub-leading jet and the missing transverse momentum, as defined in Equation \ref{eq:mt},
is required to be greater than 200 \GeV.

\begin{eqnarray}\label{eq:mt}
M_{T}(J_2, \MET) = \sqrt{[E_{T}(J_2)-E_{T}^{miss}]^2+[{\bf p_{T}}(J_2)-{\bf E_{T}^{miss}}]^{2}}, \\ \nonumber
\end{eqnarray}

The distribution of $M_{T}(J_2, \MET{})$ is expected to have a kinematic edge
at the mass of the top quark when the jet and \MET{} originate from semileptonic decay of a top quark.
Events are characterized using the boost-corrected contransverse mass ($M_{CT}$) \cite{MCT, MCT1}. For processes
with two identical decays of heavy
particles, $\tilde{b}\rightarrow J_i \lsp{}$, the $M_{CT}$ is defined as:
\begin{eqnarray}
M_{CT}^{2}(J_1, J_2) &=& [E_{T}(J_1)+E_{T}(J_2)]^2-[{\bf p_{T}}(J_1)-{\bf p_{T}}(J_2)]^{2} \\ \nonumber
   & = & 2p_{T}(J_1)p_{T}(J_2) (1+\cos\Delta\phi(J_1,J_2)),
\end{eqnarray}

To obtain sensitivity in multiple regions across the
$(m_{\tilde{b}},m_{\lsp{}})$ plane, a total of eight exclusive
search regions are defined using $M_{CT}$ and the number of b jets ($\mathrm{N_{b-jets}}$), as
summarized in Table~\ref{tab:searchbins}.

\begin{table}[h]
\begin{center}
\caption{Definition of eight exclusive search regions.}
\begin{tabular}{|l|c|c|c|c| }
\hline
No. of b-jets     &  $M_{CT}$ & $M_{CT}$   & $M_{CT}$  & $M_{CT}$  \\
\hline
$\mathrm{N_{b-jets}} = 1$  &  $<$ 250 \GeV  & 250 - 350 \GeV & 350 - 450 \GeV & $>$ 450 \GeV   \\
$\mathrm{N_{b-jets}} = 2$  &  $<$ 250 \GeV  & 250 - 350 \GeV  & 350 - 450 \GeV & $>$ 450 \GeV  \\
\hline
\end{tabular}
\label{tab:searchbins}
\end{center}
\end{table}

\begin{figure}[htb]
  \begin{center}
    \includegraphics[width=0.49\linewidth]{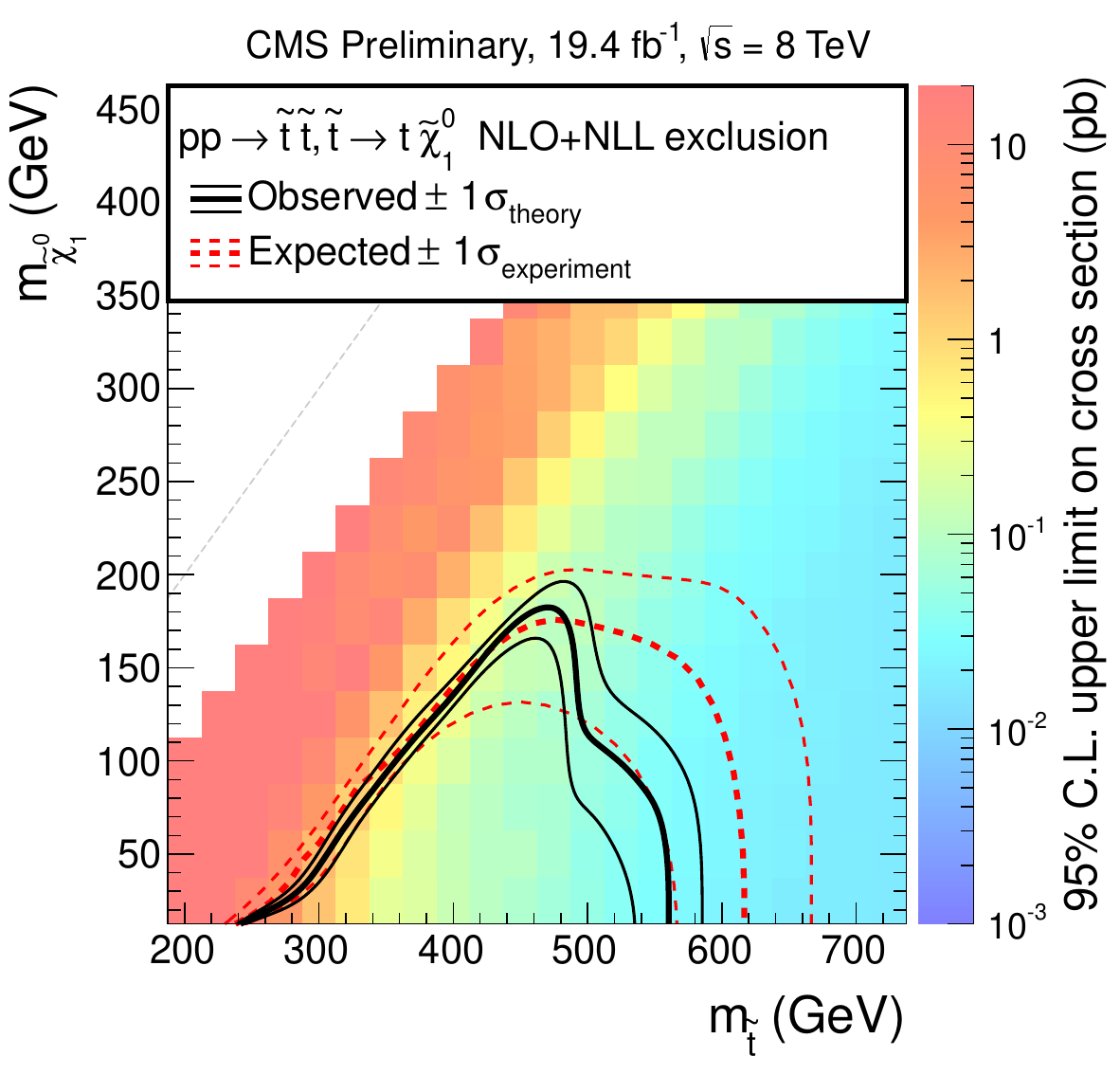}
    \includegraphics[width=0.49\linewidth]{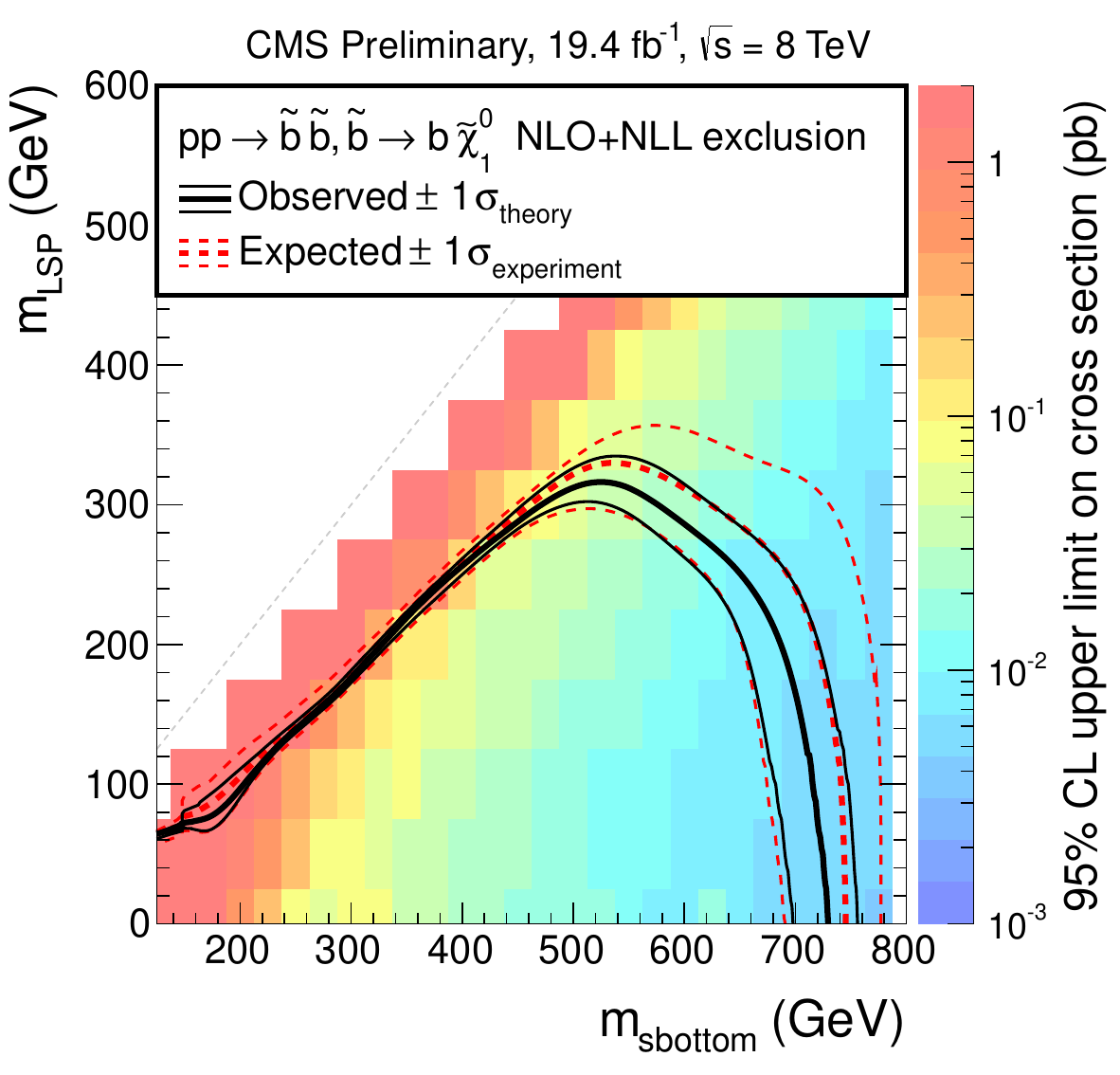}
    \caption{ Combined 95\% C.L. exclusion limits for (left) top squark pair production in the \Ttt decay channel with the all hadronic search
			  and (right) bottom squark pair production. 
			  The plot shows the expected limit as a red dashed line. The observed limit is shown as a
			  black solid line. The dashed red (solid black) lines represent the expected (observed) exclusion contours at 95\% CL.
 			  The total experimental uncertainty is shown around the expected limit contour
			  as dashed red lines, and the theoretical uncertainty is shown around the observed limit contour as thin black lines.
	    \label{fig:limitsT2ttT2bb}}
  \end{center}
\end{figure}

The major SM background contributions in this search arise from
$\znunubr$+jets, an irreducible background; $\ttbar$, single top and $\wlnubr$ + jets events, where
a W decays to  e or $\mu$ (directly or via a $\tau$ decay) which survives
the lepton veto because it is misidentified, nonisolated or is out of kinematic acceptance,
 or a $\tau$ that decays
hadronically and is reconstructed as a jet; and QCD multijet events where semileptonic
decays of b-quark jets or mismeasurements of jets can
result in large missing transverse momentum.
The full details of the event selection with analysis procedure can be found in Ref.~\cite{CMS:2014nia}.
The observed data yields are consistent with SM expectations and
subsequently limits are placed on the signal model of scalar bottom quark pair production show in Fig.~\ref{fig:limitsT2ttT2bb} (right).

\section{Search for top-squark pair production with Higgs and Z bosons} 

Searches for direct top-squark production at the LHC have focused mainly
on the simplest scenario, in which only the lighter top-squark
mass eigenstate, $\stone$, is accessible at current collision energies. 
In these searches, the top-squark decay modes
considered are those to a top quark and a neutralino,
$\stone\to \cPqt\PSGczDo \to \cPqb \PW \PSGczDo$, or
to a bottom quark and a chargino, $\stone\to \cPqb \PSGcpDo \to \cPqb \PW \PSGczDo$.
These two decay modes are expected to have large branching fractions
if kinematically allowed.
The lightest neutralino, $\PSGczDo$, is the lightest
SUSY particle (LSP) in the R-parity conserving models
considered; the experimental signature of such a particle is
missing transverse energy ($\MET$).

The sensitivity of searches for direct top-squark pair production is, however, significantly reduced
in the $\stone\to \cPqt\PSGczDo$ decay mode for the region of SUSY parameter space in which
 $m_{\stone} - m_{\PSGczDo} \simeq m_{\cPqt}$.  For example, in Ref.~\cite{Chatrchyan:2013xna},
the region $|m_{\stone} - m_{\PSGczDo} - m_{\cPqt}| \lesssim 20$\GeV is unexplored.
In this region, the momentum of the daughter neutralino in the rest frame of the decaying
$\stone$ is small, and it is exactly zero in the limit $ m_{\stone} - m_{\PSGczDo} = m_{\cPqt}$. As a result, the $\MET$
from the vector sum of the transverse momenta of the two neutralinos is typically also small in the laboratory frame.
It then becomes difficult to distinguish kinematically between
$\stone$ pair production and the dominant background,
which arises from $\ttbar$ production.
This region of phase space can be explored using events with topologies that are distinct from the $\ttbar$ background.
An example is gluino pair production where each gluino decays to a top squark and a top quark, giving rise to a
signature with four top quarks in the final state~\cite{Chatrchyan:2013iqa,Chatrchyan:2013wxa}.

This search targets the region of phase space where
$m_{\stone} - m_{\PSGczDo} \simeq m_{\cPqt}$ by focusing on signatures
of $\ttbar\PH\PH$, $\ttbar\PH\cPZ$, and
$\ttbar\cPZ\cPZ$ with $\MET$. These final states can
arise from the pair production of the heavier top-squark mass eigenstate
$\sttwo$. There are two non-degenerate top-squark mass eigenstates
($\sttwo$ and $\stone$) due to the mixing of the SUSY partners $\stL$ and $\stR$ of the
right- and left-handed top quarks. The $\sttwo$
decays to $\stone$ and an $\PH$ or $\cPZ$ boson,
and the $\stone$ is subsequently assumed to decay to $\cPqt\PSGczDo$, as
shown in the bottom row of Fig.~\ref{fig:SigDiagram}. Other decay modes such as $\stone\to
\cPqb\PSGcpDo \to \cPqb\PW\PSGczDo $ are largely covered
for $m_{\stone} - m_{\PSGczDo} \simeq m_{\cPqt}$ by existing analyses~\cite{Chatrchyan:2013xna}.
The final states pursued in this search can arise in other scenarios, such as $\stone\to
\cPqt\PSGczDt$, with $\PSGczDt \to \PH
\PSGczDo$ or $\PSGczDt \to \cPZ
\PSGczDo$. The search is also sensitive to a range of models in
which the LSP is a gravitino.
The relative branching fractions for modes with the $\PH$ and $\cPZ$ bosons are
model dependent, so it is useful to search for both decay modes simultaneously.
In the signal model considered, $\sttwo$ is assumed always to decay to
$\stone$ in association with an $\PH$ or $\cPZ$ boson,
such that the sum of the two branching fractions is
$\mathcal{B}(\sttwo\to \PH \stone) + \mathcal{B}(\sttwo\to \cPZ
\stone) = 100$\%. Other possible decay modes are $\sttwo \to
\cPqt\PSGczDo$ and $\sttwo \to \cPqb \PSGcpDo$. 

\begin{figure}[htbp]
\centering
\includegraphics[width=0.49\textwidth]{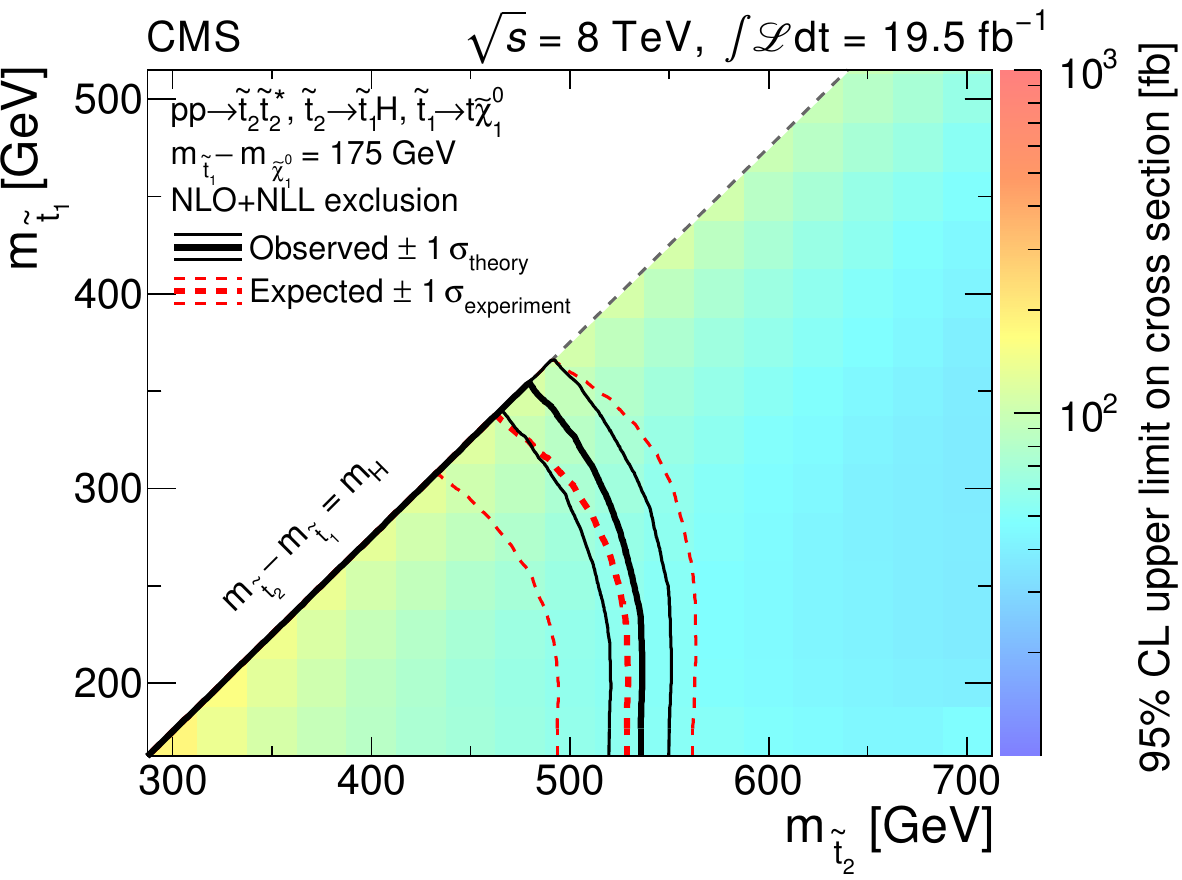}
\includegraphics[width=0.49\textwidth]{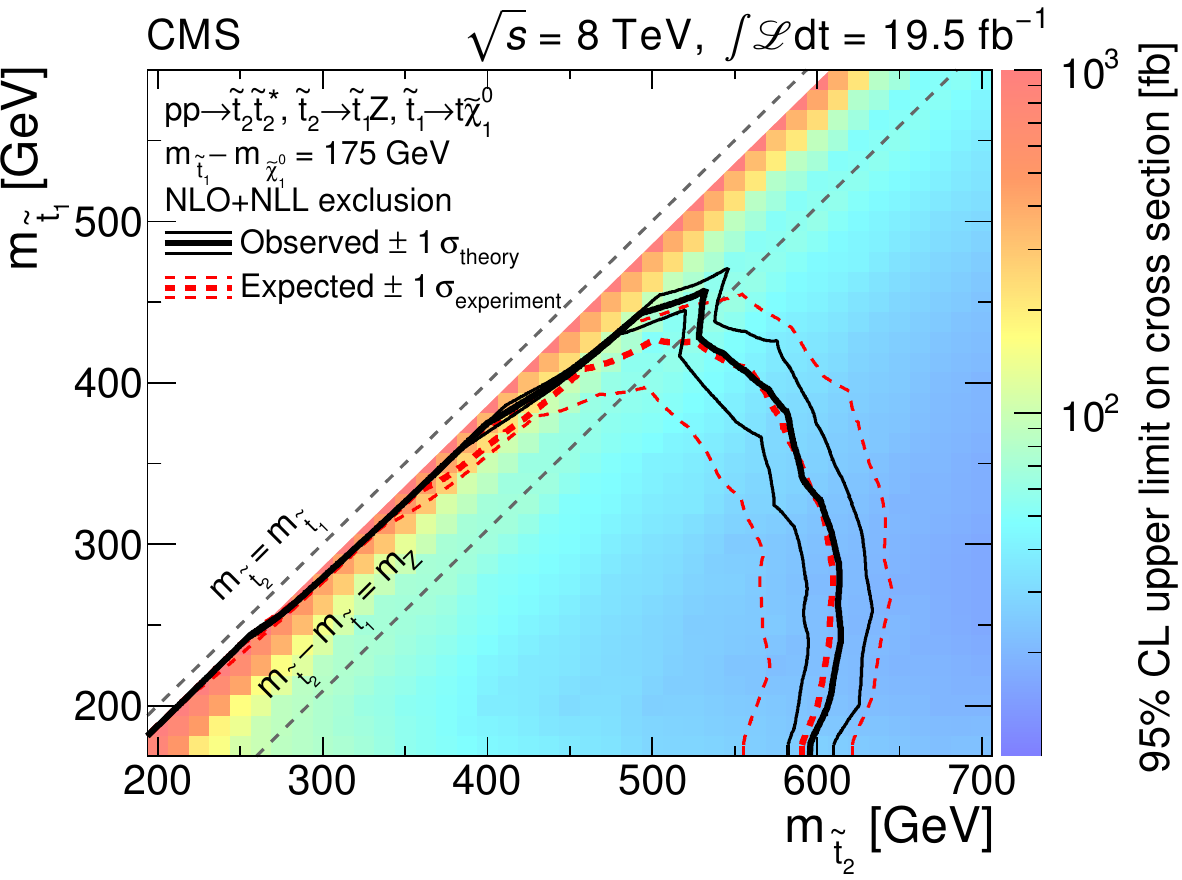}
\caption{
  Interpretation of the results in SUSY simplified model
  parameter space, $m_{\stone}$ \vs $m_{\sttwo}$, with the
  neutralino mass constrained by the relation
  $m_{\stone} - m_{\PSGczDo}  = 175\GeV$.
  The shaded maps show the upper
  limit (95\% CL) on the cross section times branching fraction
  at each point in the $m_{\stone}$ \vs~$m_{\sttwo}$ plane for the
  process $\Pp\Pp \to \sttwo\sttwo^{*}$,
  with $\sttwo\to \PH \stone$, $\stone\to
  \cPqt\PSGczDo$ and $\PSQtDt\to \cPZ\PSQtDo$,
  $\stone\to \cPqt\PSGczDo$. In these plots,
  the results from all channels are combined. The excluded region
  in the $m_{\stone}$ \vs $m_{\sttwo}$ parameter space is obtained
  by comparing the cross section times branching fraction upper
  limit at each model point with the corresponding NLO+NLL
  cross section for the process, assuming that (a)~$\mathcal{B}(\sttwo\to \PH \stone)=100\%$ or (b)~that $\mathcal{B}(\sttwo\to \cPZ \stone)=100\%$.
  \label{fig:interp_HH_and_ZZ}}
\end{figure}

The four main search channels contain either exactly
one-lepton, two leptons with opposite-sign (OS) charge and no other
leptons, two leptons with same-sign (SS) charge and no other leptons,
or at least three leptons (3 $\ell$).
The channels with one-lepton or two OS leptons require at least three
$\cPqb$ jets, while the channels with two SS leptons or 3 $\ell$ require
at least one $\cPqb$ jet.
The major SM background contribution in this search arise from $\ttbar$ pair production,
which has two $\cPqb$ quarks and either one-lepton or two OS leptons from
the $\ttbar\to \ell \nu \qqbar \bbbar$ or
$\ttbar\to \ell \nu \ell \nu \bbbar$ decay modes, where
$\cPq$ denotes a quark jet. The
sensitivity to the signal arises both from events with additional
$\cPqb$ quarks in the final state (mainly from
$\PH\to\bbbar$), and from events with additional leptons from $\PH$ or $\cPZ$ boson decays.
The full details of the event selection with analysis procedure can be found in Ref.~\cite{Chachatryan:2014doa}. 
The observed data yields are consistent with SM expectations
and subsequently limits are placed on top quark pair production (see Fig.~\ref{fig:interp_HH_and_ZZ}).

\section{Summary} 

CMS has carried out several searches for superpartners of third generation fermions with an
integrated luminosity of 19.5 fb$^{-1}$. No excess in data with respect to the SM expectation has been observed so far. 
However, searches in large regions of the parameter space for ¿natural¿ SUSY are still
in progress. These searches are challenging due to similarity with the $\ttbar$ final state for low stop masses, and due to the low cross sections for higher
stop mass values. Subsequently, new results obtained at $\sqrt{s}=13~$\TeV\ in 2015 will be looked upon with eager anticipation.

\bigskip 

\bibliography{}

\end{document}